\UseRawInputEncoding
\documentclass[onecolumn]{aastex63}
\usepackage{amsthm, amsmath, amssymb,multirow}
\begin{document}
\newcommand{\tabincell}[2]{\begin{tabular}{@{}#1@{}}#2\end{tabular}}  
\newcommand{\ratepl}{$4.9 ^{+1.9}_{-1.4} \times 10^4$}
\newcommand{\ratebpl}{$4.6 ^{+1.7}_{-1.3} \times 10^4$}
\newcommand{\plslope}{$0.80 \pm 0.16$}
\newcommand{\plc}{$4.78 \pm 0.27$}
\newcommand{\bplslopef}{$0.48 \pm 0.28$}
\newcommand{\bplslopes}{$2.11 \pm 1.27$}
\newcommand{\bpllbreak}{$45.32 \pm 0.55$}
\newcommand{\bplc}{$3.47 \pm 0.65$}

\title{Luminosity function and event rate density of \textit{XMM-Newton}-selected supernova shock-breakout candidates}

\author{Hui Sun}
\affiliation{Key Laboratory of Space Astronomy and Technology, National Astronomical Observatories, Chinese Academy of Sciences, Beijing 100012, China}

\author{He-Yang Liu}
\affiliation{Key Laboratory of Space Astronomy and Technology, National Astronomical Observatories, Chinese Academy of Sciences, Beijing 100012, China}

\author{Hai-Wu Pan}
\affiliation{Key Laboratory of Space Astronomy and Technology, National Astronomical Observatories, Chinese Academy of Sciences, Beijing 100012, China}
\affiliation{University of Chinese Academy of Sciences, School of Astronomy and Space Science, Beijing 100049, China}

\author{Zhu Liu}
\affiliation{Key Laboratory of Space Astronomy and Technology, National Astronomical Observatories, Chinese Academy of Sciences, Beijing 100012, China}
\affiliation{Max Planck Institute for Extraterrestrial Physics, Giessenbachstrasse 1, 85748 Garching, Germany}

\author{Dennis Alp}
\affiliation{Department of Physics, KTH Royal Institute of Technology, and The Oskar Klein Centre, SE-10691 Stockholm, Sweden}

\author{Jingwei Hu}
\affiliation{Key Laboratory of Space Astronomy and Technology, National Astronomical Observatories, Chinese Academy of Sciences, Beijing 100012, China}
\affiliation{University of Chinese Academy of Sciences, School of Astronomy and Space Science, Beijing 100049, China}

\author{Zhuo Li}
\affiliation{Department of Astronomy, School of Physics, Peking University, Beijing 100871, China}
\affiliation{Kavli Institute for Astronomy and Astrophysics, Peking University, Beijing 100871, China}

\author{Bing Zhang}
\affiliation{Nevada Center for Astrophysics, University of Nevada, Las Vegas, NV 89154, USA}
\affiliation{Department of Physics and Astronomy, University of Nevada, Las Vegas, NV 89154, USA}

\author{Weimin Yuan}
\affiliation{Key Laboratory of Space Astronomy and Technology, National Astronomical Observatories, Chinese Academy of Sciences, Beijing 100012, China}
\affiliation{University of Chinese Academy of Sciences, School of Astronomy and Space Science, Beijing 100049, China}

\correspondingauthor{Hui Sun}
\email{hsun@nao.cas.cn}

\begin{abstract}
A dozen X-ray supernova shock breakout (SN SBO) candidates were reported recently based on \textit{XMM-Newton} archival data, which increased the X-ray selected SN SBO sample by an order of magnitude. Assuming they are genuine SN SBOs, we study the luminosity function (LF) by improving upon the method used in our previous work. The light curves and the spectra of the candidates were used to derive the maximum volume within which these objects could be detected with \textit{XMM-Newton} by simulation. The results show that the SN SBO LF can be described by either a broken power law (BPL) with indices (at the 68$\%$ confidence level) of \bplslopef{} and \bplslopes{} before and after the break luminosity at $\log (L_b/\rm erg\,s^{-1})=$ \bpllbreak{} or a single power law (SPL) with index of \plslope{}. The local event rate densities of SN SBOs above $5\times 10^{42}$ $\rm erg\,s^{-1}$ are consistent for two models, i.e., \ratebpl{}  and \ratepl{}  $\rm Gpc^{-3}\,yr^{-1}$ for BPL and 
SPL models, respectively. The number of fast X-ray transients of SN SBO origin can be significantly increased by the wide-field X-ray telescopes such as the \textit{Einstein Probe}. 
\end{abstract}


\section{Introduction}
The supernova shock breakout (SN SBO) is the very first electromagnetic radiation of core-collapse supernova (CCSN, \citealt{falk1978, klein1978}). The radiation-mediated shock generated inside the star, either non-relativistic or relativistic, advances to the outer edge. A short X-ray and ultraviolet (UV) burst, lasting tens to thousands of seconds, is emitted at the breakout time when the optical depth of the shock equals $c/v$, where $v$ is the velocity of the shock and $c$ is the speed of light. The temporal and spectral information of SN SBOs can be used to diagnose the physical properties of the progenitor star and the shock physics (see details in chap. 5 in the \citealt{waxmankatz2017}). Gamma-ray emissions are expected when the velocity of the shock reaches the relativistic regime \citep{colgate1974, nakar2012, nakar2015, ito2020}, leading to the hypothesis of an SBO origin of the low-luminosity gamma-ray bursts (LL-GRBs; \citealt{waxman2007, nakar2012}).

Detecting fast X-ray transients with an SN SBO origin efficiently requires telescopes with a large field of view. Before the shock breaks out of the exploding star, only neutrinos and gravitational waves can escape the star. Limited by the field of view of the existing facilities and the low sensitivity of multi-messenger triggers, only two SN SBO candidates have been reported prior to the \textit{XMM-Newton} sample, i.e., XRO 080109/SN2008D  and GRB 060218/ SN 2006aj.  The former was serendipitously discovered by the {\em Swift} X-ray telescope (XRT, \citealt{soderberg2008}). The Type Ibc supernova (SN Ibc) SN 2008D was spectroscopically confirmed one day later. The SBO of SN 2008D lasted several hundred seconds and was interpreted as an SN SBO of a Wolf-Rayet (WR) star progenitor. The other SN SBO transient was identified through thermal component in the spectral analysis of GRB 060218  \citep{campana2006}.  The subsequent detection of a Type Ic supernova SN 2006aj in the optical band helped to verify the SN SBO origin. The SN SBO event in GRB 060218 can be attributed to a relativistic jet breaking out from a collapsing blue supergiant (\citealt{waxman2007}; but see \citealt{irwin2016}). Based on the detections of these two events, the progenitor properties of SN SBO, the event rate density and the luminosity function are poorly constrained. A rough event rate density of $(0.7-4.3) \times 10^4$ $\rm Gpc^{-3}\,yr^{-1}$ was estimated for SN SBOs with the luminosity similar to XRO 080109 \citep{sun2015}. The luminosity function was even less constrained, with a rough power law fit of $L^{-2}$ derived from the joint luminosity function of SN SBO and LL-GRB events \citep{sun2015}. 

Owing to their smooth light curves and accompanying supernovae in optical observations, other LL-GRBs (other than GRB 060218) have also been regarded as originating from relativistic shock breakouts \citep{wang2007, nakar2012}. In comparison with high-luminosity GRBs (HL-GRBs), LL-GRBs have displayed some distinctive characteristics including much lower-luminosity, higher local event rate density, softer peak energies, and the high observed rate in the nearby universe \citep{liang2007, virgili2009}. They also do not straightforwardly follow the extrapolation of the HL-GRB luminosity function \citep{virgili2009, sun2015}. However, it is not certain that all LL-GRBs are related to relativistic SBOs. Some LL-GRBs showed rapid variability which may be accounted for by low-luminosity jets \citep{zhang2012}. A plausible method to investigate the SBO origin of LL-GRBs is to study the LF and compare with that of nonrelativistic SN SBOs. However, there is large uncertainty in previous investigations due to the limited number of the SN SBO sample.
 
A recent work reported 12 fast X-ray transients from a search of the entire \textit{XMM-Newton} archive (\citealt{alplarsson2020}, hereafter AL20) which are new SN SBO candidates. The light curves last from 30s to ten thousand seconds. The temperatures of the blackbody fit to the spectra are within the range of $0.1 - 1$ $\rm keV$. Interpreted as SN SBOs, one can infer a diversity of the progenitors, i.e. from red supergiant stars to blue supergiant stars and to WR stars. The increased size of the SN SBO sample enables an investigation of the LF and event rate density.

In time domain astronomy, it is the number density of events per unit time, or the event rate density, that is of interest.
To derive them, one needs to compute the maximum volume ($V_{\rm max}$), or the maximum distance/redshift ($z_{\rm max}$),  that a source with a given (peak) luminosity could be detected.
The $z_{\rm max}$ depends on the detectability of the faintest possible source of that type of transient for that telescope.  
The number density is then divided by the total time span during which this maximum volume is monitored.
There are a number of factors one needs to consider to derive the detectability for fast transients, which differ from the case for persistent sources.
By fast transients we mean transients that show drastic flux changes on timescales shorter than the typical exposures of (undisrupted) observations of a telescope pointed to a certain sky region defined by its FoV (such as the SN SBO sample detected in \textit{XMM-Newton} observations).
In this case the length of the exposure time has no effect on the detection sensitivity for a fast transient (different from the case of persistent sources), but only increases the probability of its detection by contributing to the total time span of monitoring.
In addition to the peak flux, the detectability of faint sources depends also on the timescale and the shape of the light curve, the source spectrum and its variation, as well as the absorption column density for soft X-ray sources like SN SBOs.

In this paper, we study the luminosity function and event rate density of SN SBOs by improving on our previous formalism \citep{sun2015} based on the $V_{\rm max}$ method \citep{schmidt1968} and using the new \textit{XMM-Newton} data. The major improvement is that maximum volume  $V_{\rm max}$ is determined from simulations of more realistic detection of transient sources using their observed properties and the data sets of the real observations. 
We assume the AL20 \textit{XMM-Newton} SBO sample to be representative of the X-ray SN SBOs population. 
We also investigate the connection of SN SBOs and LL-GRBs. We present a $\log N- \log S$ distribution of these sources and discuss their detection prospects with future missions. 
These fast X-ray transients of an SN SBO origin are promising targets for future wide-field X-ray telescopes such as the \textit{Einstein Probe} (EP).

The article is organized as follows. The methodology for deriving the LF and event rate density for fast X-ray transients, the SN SBOs sample and the estimation of $z_{\rm max}$ are presented in Section \ref{sec:MethodSample}. The results are presented in Section \ref{sec:Results}. The discussions and conclusions are provided in Section \ref{sec:DC}. 
Throughout the paper, the newly released concordance cosmological parameters from the Planck Collaboration, i.e., $H_{0} = 67.4$, $\rm km\,s^{-1}\, Mpc^{-1}$, $\Omega_m  = 0.315$, and $\Omega_\Lambda = 0.685$, are adopted \citep{planck2020}. 

\section{Methodology and sample} \label{sec:MethodSample}
\subsection{Luminosity function of fast X-ray transients} \label{sec:method}
We follow the formalism in \cite{sun2015} to derive the luminosity function for fast transients that is based on the $1/V_{\rm max}$ method \citep{schmidt1968}, where $V_{\rm max}$ is the maximum volume within which a source can be detected for a specific telescope or survey.
The method is further improved in this work by deriving $V_{\rm max}$ from simulations of more realistic detection of transient sources using the data sets of the real observations. 
We define the redshift-dependent luminosity function $\phi(L, z)$ as the number density of transient events per unit volume, per unit time, per decade of peak luminosity, i.e. in units of $\rm Gpc^{-3}\,yr^{-1}\,dex^{-1}$. In this work, we assume that the cosmological evolution is due to the evolution of the number density and that the shape of LF (luminosity and index) does not evolve with redshift.\footnote{This assumption is applicable for the SN SBO sample in this work, since the majority of SN SBOs have redshift below 1. In case when the fast X-ray transients have a wider redshift distribution, the method can be used to derive the LF at each redshift bin.} Therefore, $\phi(L, z)$ can be written as
\begin{equation}
\phi(L, z) = f(z) \Phi(L),
\end{equation}
where the nondimensional factor $f(z)$ measures the $z$-dependence of the number density. It is defined as unity at the local universe, i.e., $f(z=0)=1$. And $\Phi(L)$ is the z-independent LF which can be taken as the local LF since it is denoted by $\phi(L | z=0) = \Phi(L)$.

The event rate density at redshift $z$ above a minimum luminosity $L_{\rm min}$, in units of $\rm Gpc^{-3}\,yr^{-1}$, is 
\begin{equation}
\begin{split}
\rho(z) &= \int_{\log L_{\rm min}}^{\infty} \phi(L,z)d \log L \\
& = f(z) \int_{\log L_{\rm min}}^{\infty} \Phi(L)d \log L \\
& = f(z)\rho_0(> L_{\rm min}),
\end{split}
\end{equation}
where  $\rho_0$ is defined as the local event rate density, i.e.,
\begin{equation}
\rho_0(> L_{\rm min}) = \int_{\log L_{\rm min}}^{\infty} \Phi(L)  d\log L,
\label{eq:rho}
\end{equation}
and $\log $ denotes $\log _{10}$ here and throughout the paper.  

For a given survey, the expected detectable number of transients in the luminosity range from $\log L$ to $\log L + d\log L$  and in the redshift range from $z$ to $z+dz$ can be expressed as
\begin{equation}
dN(L,z)= \frac{\Omega}{4\pi}\cdot dV(z) \cdot\phi(L,z)d \log L  \cdot T \cdot  \frac{1}{1+z},
\label{eq:dN(L,z)}
\end{equation} 
where $\Omega$ is field of view (FoV) of the telescope/survey, the volume element $dV$ over the 4$\pi$ solid angle is given by $dV = \frac{dV}{dz} \cdot dz$ and $T$ is the time duration that this volume element is monitored. The $\frac{1}{1+z}$ factor corrects for the effect of time dilation, as the observation time $T$ is measured in the local frame of the observer.
In the standard $\Lambda$CDM cosmology, the comoving volume is given by
\begin{equation}
\frac{dV(z)}{dz}=\frac{c}{H_0}\frac{4\pi D_L(z)^2}{(1+z)^2[\Omega_M(1+z)^3+\Omega_{\Lambda}]^{1/2}},
\end{equation} 
where $D_L(z)$ is the luminosity distance at the corresponding redshift $z$.

The $f(z)$ factor is model dependent. By assuming that core-collapse supernovae trace the star formation history (SFH), we adopt the analytical formula from \cite{yuksel2008},
\begin{equation}
f(z)=\left( (1+z )^{3.4\eta}+\left( \frac{1+z}{5000}\right) ^{-0.3\eta}+\left( \frac{1+z}{9}\right) ^{-3.5\eta} \right) ^{\frac{1}{\eta}},
\label{eq:sfh}
\end{equation}
where $\eta=-10$. At $z < 4$, this function measures the SFH from the UV and far-IR galaxy data \citep{hopkinsbeacom2006, li2008}. At redshifts beyond $z > 4$, the SFH is enhanced taking into account of the high-redshift GRBs by the Swift data, which exceeds the SFH directly inferred from the galaxy observation as far as $z \sim 7$.

In practice, the survey may be composed of a number of observations. Equation \ref{eq:dN(L,z)} can be written as 
\begin{equation}
dN(L,z) = \sum_{i=1} ^{N_{\rm obs}}dN(L,z)_{i} ,
\end{equation}
where $dN(L,z)_{i}$ is the expected number for $i$th observation and $N_{\rm obs}$ is the total number of the observations.
The detectable source number with $L$ is the integral from $z=0$ to $z_{\rm max,i}$ for the $i$th observation, i.e.,
\begin{equation}
dN(L)_{i} =  \Phi(L) d \log L \cdot T_{i} \cdot \int_{0}^{z_{\rm max,i}} \frac{\Omega}{4\pi} \cdot \frac{f(z)}{1+z} \frac{dV(z)}{dz} dz,
\label{eq:dNi}
\end{equation}
where $z_{\rm max,i}$ is the maximum redshift at which the source can be detected by the observation and $T_i$ is the exposure time of the observation. 

Therefore the total detectable number of sources with $L$ is the sum over all the observations,
\begin{equation}
dN(L)= \Phi(L)d\log L \cdot  \sum_{i=1} ^{N_{\rm obs}} T_i \cdot V'_{\rm max, i},
\end{equation} 
where $V'_{\rm max, i}$ is the effective maximum volume that is monitored by the $i$th observation weighted by the density evolution and time dilation, i.e.,
\begin{equation}
V'_{\rm max, i} = \int_{0}^{z_{\rm max,i}} \frac{\Omega}{4\pi} \cdot \frac{f(z)}{1+z} \frac{dV(z)}{dz} dz.
\label{eq:V'max}
\end{equation}
 
We denote the parameter $\mathcal{M}$ as the total monitored volume-time (MVT) of all the observations, i.e.,
\begin{equation}
\mathcal{M} = \sum_{i=1} ^{N_{\rm obs}} T_i \cdot V'_{\rm max, i}.
\label{eq:TV}
\end{equation}

For a homogeneous population, i.e., sources differ only in luminosity and redshift, the maximum redshift depends mainly on the source peak luminosity. If the observed sample comprises objects that differ in other quantities, such as light curves, spectral shape (e.g. blackbody temperatures), etc., they can be considered as being drawn from a mixture of several populations.
The luminosity function $\Phi(L)$ is the sum of the individual contributions by each source type ($j$) in the range from $\log L$ to $\log L + d\log L$ (with total $\Delta N_{L}$ sources), i.e.,
\begin{equation}
\begin{split}
\Phi(L)d \log L &= \sum_{j=1}^{\Delta N_{L}} \left[ \Phi(L)d \log L \right]_j \\
&= \sum_{j=1}^{\Delta N_{L}} \frac{dN(L)_{j}}{\left[ \sum_{i=1} ^{N_{\rm obs}} T_i \cdot V'_{\rm max, i}\right]}_j \\
& = \sum_{j=1}^{\Delta N_{L}} \frac{1}{\mathcal{M}}_j.
\end{split}
\label{eq:lf}
\end{equation}

\subsection{SN SBO sample}
Our work is based on the assumption that the search method in AL20 resulted in a complete sample of SN SBOs from the \textit{XMM-Newton} data used. 
These include
all publicly available archival data as of 2019 November in HEASARC and from 3XMM-DR8. The total good exposure time\footnote{Sensitive searches were only possible for exposures taken in a large FoV imaging mode (Extended Full Frame, Full Frame, or Large Window) and during low background intervals. These values are for the pn camera, but those for the MOS cameras are similar.} is 163 Ms (5.2 yr). 
Of this, only 123 Ms (3.9 yr) is outside of the Galactic plane ($|b| > 15^\circ$). 
The effective field of view of \textit{XMM-Newton} observations is taken as 0.2 square degrees. 

\begin{figure}
\centering
\includegraphics[scale=0.5]{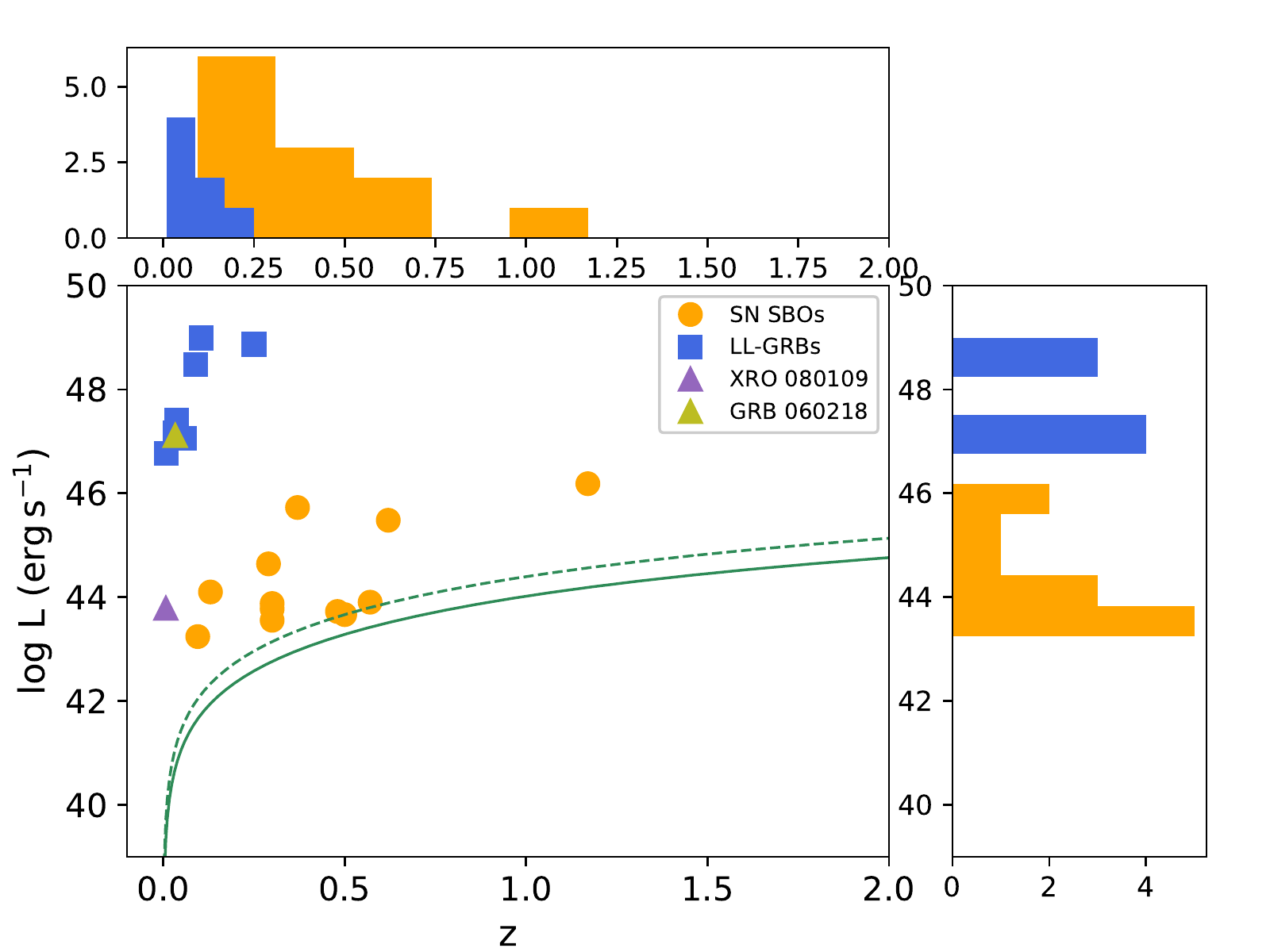} 
\caption{The peak bolometric luminosity and redshift distribution of SN SBO candidates and LL-GRBs (\citealt{sun2015}). The green dashed line assumes the minimum of the peak flux of the \textit{XMM-Newton} sample and the solid line represent how the \textit{XMM-Newton} EPIC threshold varies as a function of redshift for a 1000$s$-exposure in $0.5 - 2$ $\rm keV$ \citep{watson2001}.}
\label{fig:L}
\end{figure} 

The total sample inferred from AL20 is shown in Table \ref{tab:sample}. The silver sample differs with the golden one due to the insufficient redshift estimate or possible Galactic foreground contamination.
Following AL20, a pseudo-redshift of 0.3 is assumed for the candidates without redshift measurements. The SN SBO peak bolometric luminosities and the temperatures derived assuming the blackbody model are shown in Table \ref{tab:sample}. In Figure \ref{fig:L}, we show the distribution of the peak bolometric luminosities and redshifts for \textit{XMM-Newton} SN SBO candidates and LL-GRBs. The majority of the LL-GRB sample are summarized in \cite{sun2015}, supplemented with the recently-reported event LL-GRB 171205A \citep{D'Elia2018}. The bolometric luminosities of SN SBOs normally peak in the soft X-ray band, with the blackbody temperatures ranging in $\left[ 0.1, 1\right] $ $\rm keV$. The peak bolometric luminosities of the newly-reported SN SBO candidates generally reside in the range of $\left[ 10^{42}, 10^{47}\right] $ $\rm erg\,s^{-1}$. The peak luminosity of XRO 080109/SN 2008D is located within the range of the  \textit{XMM-Newton} sample. GRB 060218/SN 2006aj has a higher luminosity than all the \textit{XMM-Newton} sample. 

For illustration purposes, we also show how the limiting luminosity corresponding to the \textit{XMM-Newton}  detection threshold varies as a function of redshift for an exposure time of $\sim 1000s$, (roughly the timescale for a typical SN SBO), in soft X-rays (corresponding to $F_{\rm th} = 2 \times 10^{-14}$ $\rm erg\,cm^{-2}\,s^{-1}$ in the range of $0.5 - 2$ $\rm keV$, \citealt{watson2001}) with the solid line. The minimum peak flux of the sample is higher than this sensitivity limit, as shown with the dashed green line ($F_{\rm peak, min} = 4.5 \times 10^{-14}$ $\rm erg\,cm^{-2}\,s^{-1}$).  All SN SBOs lie above the $F_{\rm th}$ curve and therefore the sample is generally considered as a complete SN SBO sample above this flux limit.

\begin{table*}
\caption{The \textit{XMM-Newton} SN SBO candidates sample.\label{tab:sample}}
\renewcommand{\arraystretch}{1.2}
\begin{center}
\begin{tabular}{lcccc}
\hline 
\hline
Names & $L_{p,\rm bol}$ &$ z_{\rm obs}$ & $T_{\rm bb}$ & Progenitor types\tablenotemark{a}\\ 
 & ($10^{44}$ $\rm erg\,s^{-1}$) & & ($\rm keV$) & \\
\hline 
Golden sample \tablenotemark{b}\\
\hline
XT 161028 & 3.39 & 0.29 & 0.41 &  BSG  \\
XT 151219 & 32.88 & 0.62 &0.37 & BSG \\
XT 110621 & 0.39 & 0.095 &0.42 & BSG/WR  \\
XT 030206 & 172.47 & 1.17 &0.47 & SBO from an extreme SN \\
XT 070618 &  48.78 & 0.37 &0.46 & BSG\\
XT 100424 &  0.05 & 0.13 &0.13  & RSG \\
XT 151128 &  1.83 & 0.48 &0.16  & RSG\\
\hline
Silver sample\\
\hline
XT 060207 & 6.44& 0.3*\tablenotemark{c}& 0.93 & BSG  \\
XT 050925 & 0.21& 0.3*&0.45 & BSG/WR/Galactic   \\
XT 160220 & 0.51& 0.3*&0.45 & BSG/WR/Galactic  \\
XT 140811 & 0.93& 0.57&0.32 & BSG/WR/Galactic  \\
XT 040610 &0.32& 0.5&0.68 & BSG/WR/Galactic  \\

\hline
\hline
\end{tabular} 
\tablenotetext{a}{The progenitor types of the SN SBOs are mostly uncertain.
Both BSG and WR are possible origins for the candidates labelled with BSG/WR. XT 030206 is the most energetic event and could be an SBO from an extreme SN, see section 7.5 in AL20 for detailed information. }
\tablenotetext{b}{The silver sample are those events with insufficient redshift estimate or possibly Galactic foreground contamination.}
\tablenotetext{c}{*Assumed redshifts for sources without identified host galaxies.}
\end{center}
\end{table*}

\subsection{Estimation of $z_{\rm max}$ } \label{sec:Vmax}
From Equation \ref{eq:lf}, to derive the luminosity function, one needs to quantify the total MVT for each SN SBO of the \textit{XMM-Newton} sample.
The key is to determine the maximum redshift $z_{\rm max}$ for each of the observations up to which a given source could be detected. Its value depends not only on the source property, but also on the several observational factors that impact the detection of the source, including the Galactic absorption column density ($N_{\rm H, gal}$) on the line of sight (LoS), the detector filter used, the off-axis angle of the source in the FoV， and the background. 
In this work we examine the detectability of an SN SBO by simulating observations with \textit{XMM-Newton} and analyzing the resulting data, as follows. 

For a hypothesized SN SBO with a given redshift, peak luminosity, X-ray spectrum, light curve and intrinsic absorption (if any), we simulate the detection of such a source for each of the \textit{XMM-Newton} observations (with $|b| > 15^\circ$ in the large FoV imaging mode) considered. 
First, its flux spectrum model after Galactic absorption is calculated in the observer's frame using the $N_{\rm H, gal}$ value on the LoS of that pointed observation. 
This model is then convolved with the energy response and effective area of the three EPIC detectors, pn, MOS1 and MOS2, to generate detected source counts, respectively. 
For this, an averaged effective area, weighted by off-axis angle, is calculated and adopted to account for the vignetting effect of the telescope, assuming that the source off-axis angle is randomly distributed within the FoV. 
The resultant source counts of the three cameras are co-added to form a single source light curve.  
The background count rate is randomly sampled from the \textit{XMM-Newton} observations where background flares are excluded \citep{carter2007}, and then a background light curve is generated following a Poisson distribution. 
This transient source is considered to be detected if all the following criteria are met.

$\bullet$	The signal-to-noise ratio of the detected source counts is above 5.

$\bullet$	The total net source counts are no less than $N_{\rm ph,min} = 62$, the minimum counts of the \textit{XMM-Newton} SN SBOs in the AL20 sample.

$\bullet$	The ratio of peak flux to average flux is at least above 1.9, which is the minimum of the observed sample.

The last two are adopted to ensure that a simple transient-like light curve can be constructed and identified, as in AL20. 

For each SN SBO source of the observed sample, we derive the $z_{\rm max}$ for each of the \textit{XMM-Newton} observations considered. 
We assume the source to be at a set of redshifts starting from $z=0$, and we calculate its light-curve models using the source properties taken from AL20, i.e. the peak luminosity, blackbody temperature, light curve, and intrinsic absorption column density. 
Then, each of the light-curve models is taken as input to the above simulation to examine whether the source at the assumed redshift can be detected or not, as described above, for each of the \textit{XMM-Newton} observations considered. 
The simulation is carried out for light-curve models with increasing redshift values until a $z_{\rm max}$ is reached, beyond which the source can no longer be detected. 

\begin{figure}
\centering
\includegraphics[scale=0.5]{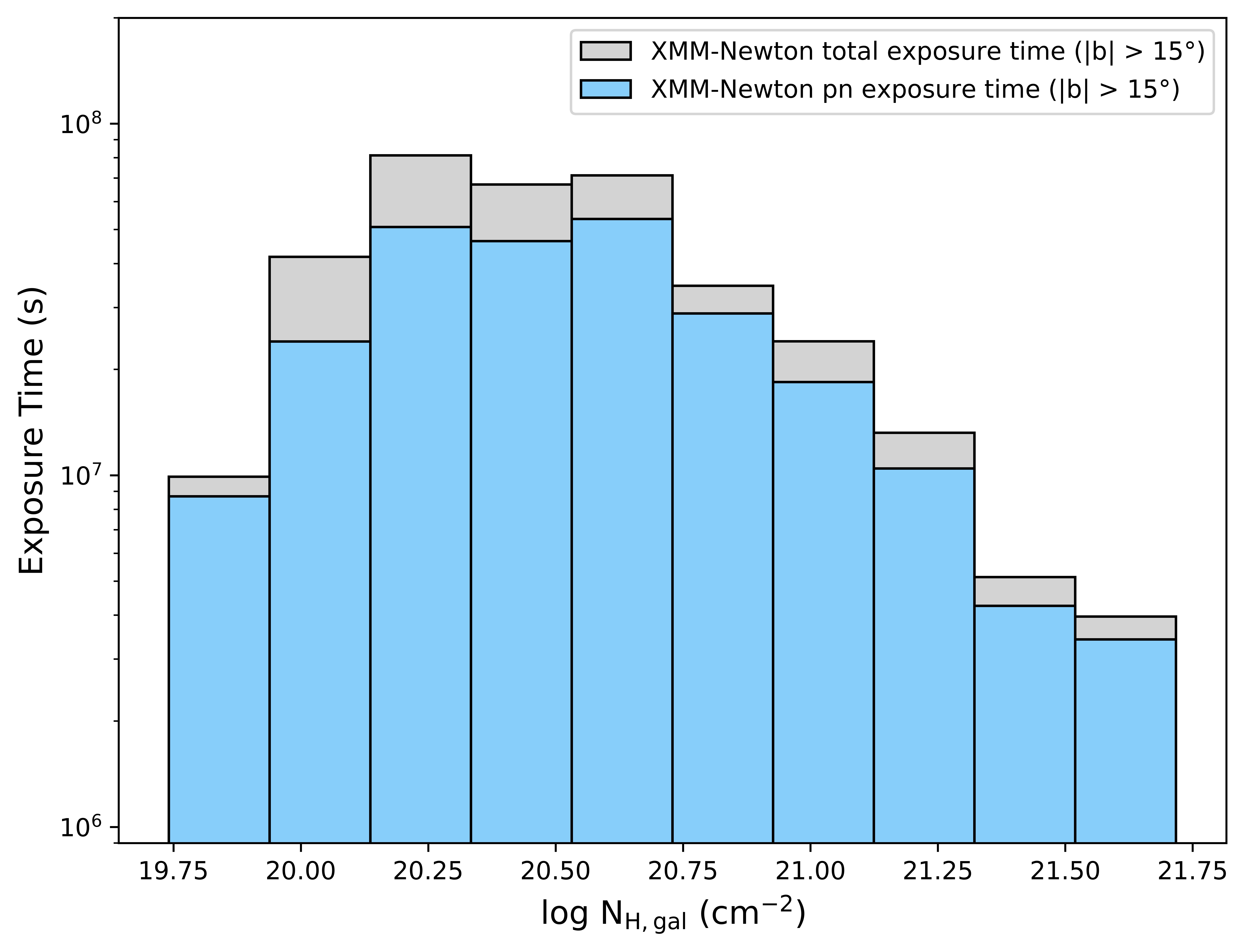}
\caption{\textit{XMM-Newton} exposure times as a distribution of the Galactic column densities. The gray histogram represents the total exposure time outside of the Galactic plane ($|b| > 15^\circ$)  and the blue histogram is the total pn exposure time at $|b| > 15^\circ$ (in the large FoV imaging mode).}
\label{fig:Nh}
\end{figure}

In this way, for a given source in the SN SBO sample, the $z_{\rm max}$ values are derived for all the observations at $|b| > 15^\circ$ during low background intervals, and the total MVT is calculated using Equations \ref{eq:V'max} and \ref{eq:TV}.
Since it is expected that a source will statistically have the same $z_{\rm max}$ values from the above simulations for observations with the same  $N_{\rm H, gal}$,  
in practice, to save computing time, we group all the observations considered into 10 bins according to their $N_{\rm H, gal}$ values in logarithm. For each $N_{\rm H, gal}$, the above process is repeated 10 times by randomly sampling the background and the average of value of $z_{\rm max}$ is taken as the final maximum redshift. In calculating the Equation \ref{eq:TV}, the exposure times are first summed up for observations within the same $N_{\rm H, gal}$ bin. Figure \ref{fig:Nh} shows the distribution of the summed exposure times in different $N_{\rm H, gal}$ bins.
Repeating the procedures above, the  parameter $\mathcal{M}$
is calculated for all the 12 sample objects. The luminosity function $\Phi(L)$ is derived by summing up the contribution of each individual source in the same luminosity bins (Equation \ref{eq:lf}).

\section{Results} \label{sec:Results}
\subsection{Luminosity function and event rate density}

\begin{figure}
\centering
\includegraphics[scale=0.6]{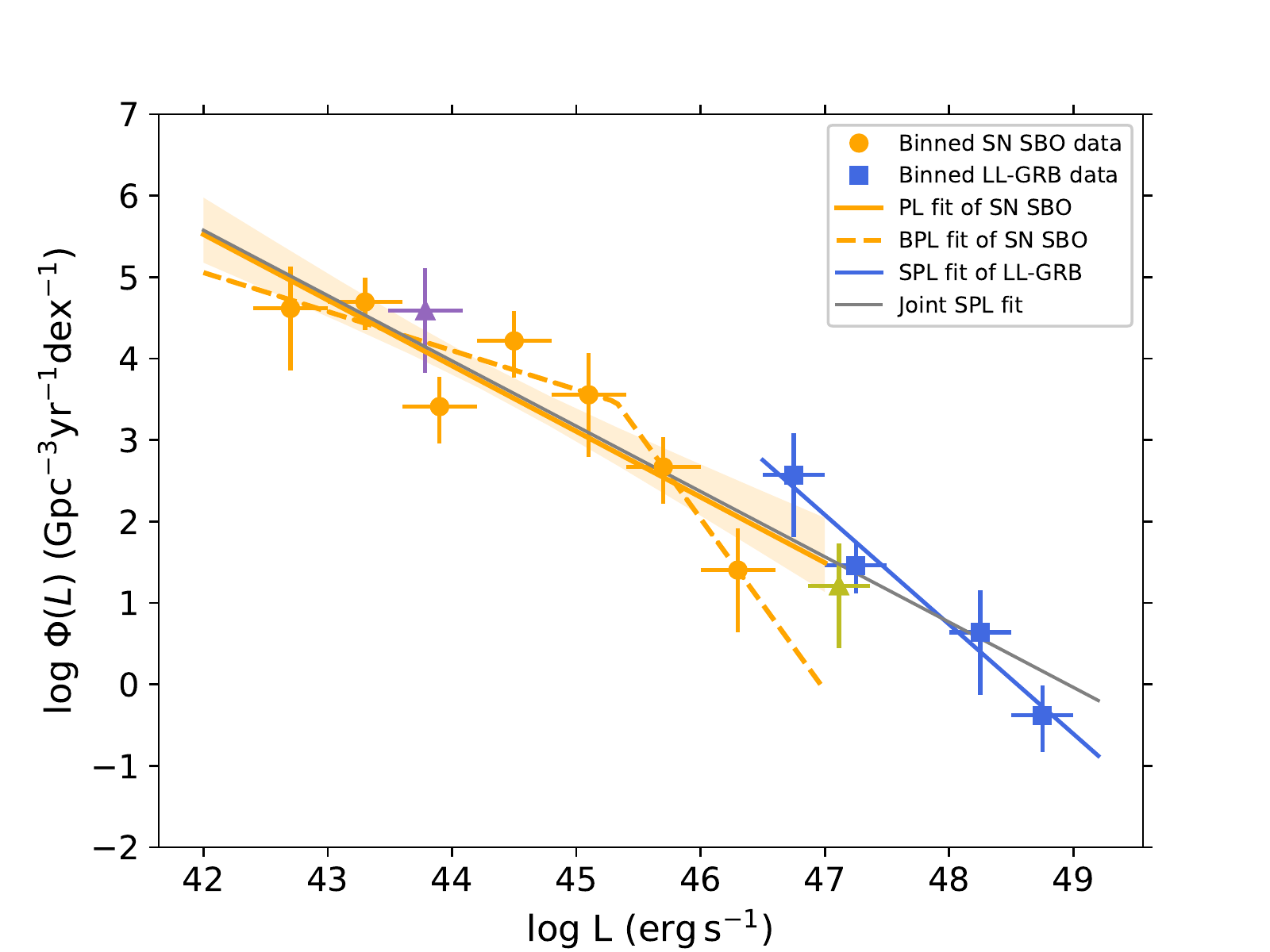} 

\caption{The luminosity function in logarithm for SN SBOs. The orange data are binned data of SN SBOs. The orange solid and dashed lines represent the SPL and BPL, respectively. The 68$\%$ confidence region of the SPL is given within the light-orange lines. The purple and olive triangles denote separately the data of XRO 080109 and GRB 060218, which are imported from \cite{sun2015}. The data of LL-GRBs (blue) and the SPL fit are also imported from \cite{sun2015}. The gray line is the joint SPL fit for both SN SBOs and LL-GRBs.}
\label{fig:lf}
\end{figure}

The luminosity function of SN SBOs derived above based on the full sample of candidates is shown in Figure \ref{fig:lf}. The horizontal error bars represent the width in logarithmic luminosity ($ \Delta \log L_{\rm SN SBO} = 0.6$).  The vertical error bars represent the 1$\sigma$ Gaussian errors calculated following \cite{gehrels1986} using the number of events in each luminosity bin. We fit the luminosity functions of SN SBOs with both a single power law (SPL) and a broken power law (BPL) using astropy curve-fit. A linear least-squares algorithm has been applied to perform the fitting. We use the Python package \textit{Uncertainties} to derive the standard deviation of the fitting parameters.

The fitted results are summarized in Table \ref{tab:lffit}. It is shown that the LF generally decreases with increasing luminosity, followed by a steep decay around $L\sim 10^{45}$ $\rm erg\,s^{-1}$, initiating a motivation of fitting with a BPL. We derive the BPL fit as
\begin{equation}
\log \Phi_{\rm BPL}(L)=\left\{
\begin{array}{rcl}
- \alpha_1 (\log L - \log L_b) +c_b  & & {\log L < \log L_b}\\
- \alpha_2 (\log L - \log L_b) +c_b   & & {\log L > \log L_b}\\
\end{array} \right.
\end{equation}
with the indices (at the 68$\%$ confidence level, same as the following) $\alpha_1 =$ \bplslopef{} and $\alpha_2 =$ \bplslopes, $c_b =$ \bplc{}, and the break luminosity $\log (L_b/\rm erg\,s^{-1})=$ \bpllbreak{}.  In this model, the local event rate density above the minimum luminosity  $L_{\rm min} = 5 \times 10^{42}$ $\rm erg\,s^{-1}$ is derived from Equation \ref{eq:rho}, i.e.,
\begin{equation}
\begin{split}
\rho_{0, \rm BPL} (> L_{\rm min} ) &= \int_{\log L_{\rm min}}^{\infty} \Phi_{\rm BPL}(L)d \log L \\
& \simeq 4.6 ^{+1.7}_{-1.3} \times 10^4 \, \rm Gpc^{-3}\,yr^{-1}.
\end{split}
\end{equation}

In general, one can also use an SPL model to fit the LF which is given as
\begin{equation}
\log \Phi_{\rm SPL}(L)=-\alpha_0 ( \log L - \log L_0) + c_0.
\end{equation}
By normalizing to $\log (L_0/\rm erg\,s^{-1}) = 43$, we obtain $\alpha_{0} =$ \plslope{}, and $c_0 =$ \plc.
The local event rate density above $L_{\rm min} = 5 \times 10^{42}$ $\rm erg\,s^{-1}$ is 
\begin{equation}
\begin{split}
\rho_{0,\rm SPL} (> L_{\rm min} ) &= \int_{\log L_{\rm min}}^{\infty} \Phi_{\rm SPL}(L)d \log L \\
& \simeq 4.9^{+1.9}_{-1.4} \times 10^4 \, \rm Gpc^{-3}\,yr^{-1},
\end{split}
\end{equation}
which is close to $\rho_{0, \rm BPL}$.  It should be noted that the LF in this work is the one for logarithmic luminosity. In case of the luminosity function in linear luminosity space, the absolute value of the index ($\beta$) should be steeper by $1$ than the index used here ($\alpha$), i.e., $ \beta = \alpha + 1$.  

To estimate the uncertainty induced by the width of bins, we also perform the fitting for the case of choosing $ \Delta \log L_{\rm SN SBO} = 0.9$. The results are also summarized in Table \ref{tab:lffit}.  Due to the small number of the sample, the best-fit values vary for different width choices. Within the uncertainties, the results from different bin widths are consistent for both BPL and SPL models. We take $ \Delta \log L_{\rm SN SBO} = 0.6$ as a representative for analysis in the following context.

The results are based on the assumption that the cosmological evolution of SN SBOs follows the description in Equation \ref{eq:sfh}. The effect of this assumption on the result should be limited since the majority of the sample is distributed at low redshift (except one candidate that has a redshift of 1.17). We also test the SFH model in \cite{li2008} and find that the differences in the results are negligible.

It should also be noted that the LF is based on the assumption that the observed 12 sample members are the representatives of all X-ray SN SBO populations. In AL20, four of the candidates, XT 050925, 160220, 140811, and 040610, are possibly misidentified Galactic foreground sources. Besides, one more event, XT 060207 is without a host redshift estimate. To estimate the uncertainty caused by the confidence of the sample being an SN SBO and the redshift measurement, we perform a further fitting for the 'Golden' sample by excluding these five events. The rest of the sample is not enough for a bin width of $ \Delta \log L_{\rm SN SBO} = 0.6$. By choosing a bin width of $ \Delta \log L_{\rm SN SBO} = 0.9$, the BPL fitting gave the power-law indices of $\alpha'_{1} = 0.00 \pm 1.01$, $\alpha'_2 = 1.59 \pm 0.40$ before and after the break luminosity at $\log (L'_b/\rm erg\,s^{-1})= 44.17 \pm 0.79$ and SPL fitting gave the slope of $\alpha'_0 = 0.95 \pm 0.21$. The results derived for the 'Golden' sample are in agreement with those for the full sample with the same width of binning, as shown in Table \ref{tab:lffit}. A larger sample is needed in the future to improve the LF measurement.

\begin{table*}
\caption{Results of luminosity function and local event rate densities. The errors are given at the 68$\%$ confidence level.\label{tab:lffit}}
\renewcommand{\arraystretch}{1.2}
\begin{center}
\begin{tabular}{c|c|c|c|c}
\hline
\hline
\multirow{2}{*}{Models} & \multirow{2}{*}{Parameters} & \multicolumn{2}{c|}{Full sample} & Golden sample\\
\cline{3-5}
 &  & $ \Delta \log L_{\rm SN SBO} = 0.6$ & $ \Delta \log L_{\rm SN SBO} = 0.9$ &  $ \Delta \log L_{\rm SN SBO} = 0.9$  \\
\hline
BPL & $\alpha_1$ & \bplslopef &  $-0.11 \pm 0.77$ & $0.00 \pm 1.01$\\

& $\alpha_2$ & \bplslopes & $1.27 \pm 0.35 $ & $1.59 \pm 0.40 $ \\
& $\log (L_b/\rm erg\,s^{-1})$ & \bpllbreak & $44.15 \pm 0.53$ & $44.17 \pm 0.79$  \\
& $c_b$ & \bplc &  $4.60 \pm 0.57$ &  $4.44 \pm 1.17$\\
& $\rho_{0, \rm BPL}$\tablenotemark{a}  & \ratebpl  & $6.2 ^{+2.4} _{-1.8} \times 10^4$ & $5.6 ^{+2.1} _{-1.6} \times 10^4$ \\
\hline
SPL & $\alpha_0$ & \plslope &  $0.75\pm 0.17 $ &  $0.95\pm 0.21$\\
& $c_0$ & \plc & $4.94 \pm 0.28$ & $4.84 \pm 0.41$ \\
& $\rho_{0, \rm SPL}$  & \ratepl  &  $5.2 ^{+2.0} _{-1.5} \times 10^4$ &  $8.0 ^{+3.0} _{-2.3} \times 10^4$\\
\hline
\hline
\end{tabular}
\tablenotetext{a}{The local event rate density, in unit of $\rm Gpc^{-3}\,yr^{-1}$, is the one above $5 \times 10^{42}$ $\rm erg\,s^{-1}$ in this table and the 1$\sigma$ Gaussian errors are derived from \cite{gehrels1986}. }
\end{center}
\end{table*}

\subsection{Comparison with LL-GRBs and other high energy transients}
It has been shown that several LL-GRBs (e.g. GRB 980425, GRB 031203, GRB 061218, GRB 100316D) may have SBO origins as they can be well explained by the relativistic breakout relation of energy, temperature, and duration (see \citealt{nakar2012} and references therein). Previously, a joint fit of XRO 080109 and LL-GRB was described by an SPL with the index of $\beta = 2$, which is similar to the LF of LL-GRB \citep{sun2015}. We investigate how the SN SBOs and LL-GRBs are distributed in the LF with the new data. In fact, LL-GRBs are detected in hard X-ray/gamma-rays and SN SBOs are detected in soft X-rays. We compare their bolometric luminosities which peak at different energy bands. The bolometric luminosities of SN SBOs are given in Table \ref{tab:sample} with the blackbody temperatures ranging in $\left[ 0.1, 1\right] \,\rm keV$. The bolometric luminosities of LL-GRBs are defined in the energy range of $1-10^4$ $\rm keV$, and derived through k-correction assuming that the spectrum follows a standard GRB Band function (\citealt{band1993,sun2015}). The data and fitted line of LL-GRB are inferred from \cite{sun2015} and improved with the event GRB 171205A.\footnote{The results of LL-GRBs with one more event are derived following the same method of \cite{sun2015}.}

Based on the new sample, we perform a new joint fit for SN SBOs and all the LL-GRBs and obtain an SPL index of $\alpha_{j} = 0.84 \pm 0.08$, similar to the previous one. The joint fit result is very close to the SPL fit of SN SBOs, as shown in Figure \ref{fig:lf}. There is also controversial case emerging as LL-GRB 120422A \citep{zhang2012}, with a duration much shorter than that predicted from the relation.\footnote{GRB 120422A is not included in our LL-GRB sample as it is regarded as engine driven origin.} It has been proposed that there is a luminosity threshold of $10^{48}$ $\rm erg\,s^{-1}$, above/below which LL-GRB may be central engine driven or relativistic shock breakout originated.
Under this assumption, we also perform another joint fit for SN SBOs and LL-GRB below $10^{48}$ $\rm erg\,s^{-1}$, the SPL fit gives an index of $\alpha'_{ j} = 0.76 \pm 0.10$, which is also similar to $\alpha_0$. From both joint fits, one can infer that a fair fraction of LL-GRBs follow the extension of SN SBO SPL LF, in agreement with their relativistic shock breakout origins.

The data of XRO 080109 and GRB 060218 that were derived independently based on the Swift detection in \cite{sun2015} are plotted in Figure \ref{fig:lf}. For comparison, their horizontal error bars are taken with the same values as those of SN SBOs and LL-GRBs.
Within the error bars, the one of XRO 080109 is consistent with the distribution of SN SBOs for both models. The one of LL-GRB 060218 is more consistent with the SPL extrapolation of SN SBOs LF. 

\begin{figure}
\centering
\includegraphics[scale=0.6]{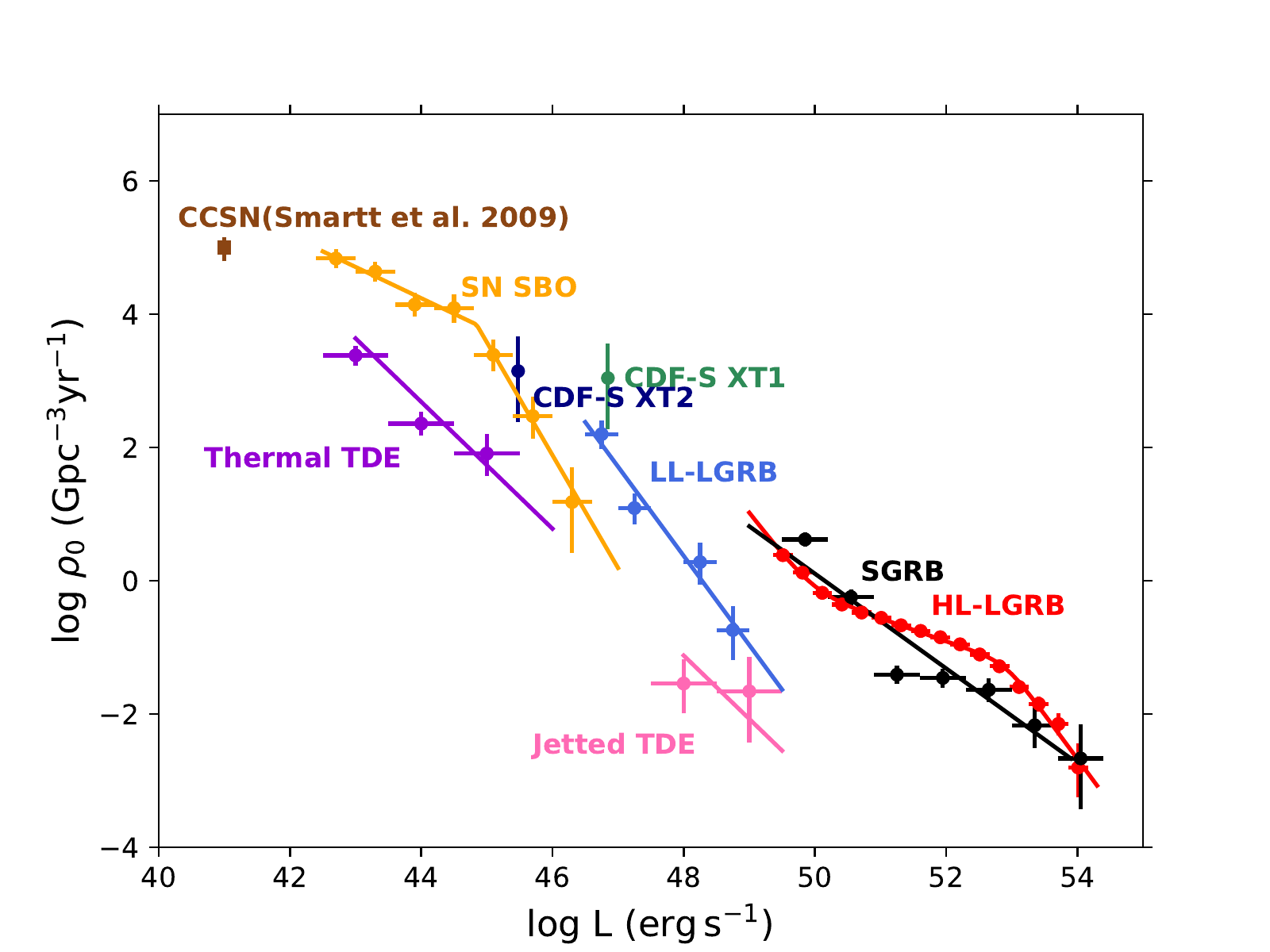} 
\caption{The cumulative local event rate density as a function of peak bolometric luminosity for various high-energy transients. The orange star denotes the SN SBO results derived in this work. The rate for CCSNe is inferred from \cite{smartt2009}, those for CDF-XT1 and XT2 are taken from \cite{sun2019}, and the rest are inferred from \cite{sun2015}. The local event rate densities above a minimum luminosity are shown in Table \ref{tab:summaryrate}. }
\label{fig:eventrate}
\end{figure}

The LFs and event rate densities of other high-energy transients, including HL-LGRBs, short GRBs (SGRBs), thermal tidal disruption events (TDEs), jetted TDEs and binary neutron star merger magnetar-powered X-ray transients, have also been investigated (e.g., \citealt{sun2015} and references therein; \citealt{sun2017, sun2019, auchettl2018, xue2019}). In Figure \ref{fig:eventrate}, we display the local event rate density as a function of the peak bolometric luminosity for SN SBOs and other high-energy transients. Similar to LL-GRBs, the bolometric luminosity of other high-energy transients are also in the range of $1-10^4$ $\rm keV$ \citep{sun2015}, except for CDF-S XT1 and XT2, whose luminosities are in the range of $0.3 - 10$ $\rm keV$ owing to the lack of knowledge of their energy spectrum.
The typical local event rate densities above the minimum luminosities are also summarized in Table \ref{tab:summaryrate}. These high-energy transients are all promising sources for future wide-field X-ray telescopes. 
With threshold luminosity spanning over $10^{42} - 10^{50}$ $\rm erg\,s^{-1}$, the local event rate densities decrease from $10^4$ to order of unity in units of $\rm Gpc^{-3}\,yr^{-1}$.
\begin{table*}
\caption{Summary of event rate densities of high-energy transients above the minimum luminosities.\label{tab:summaryrate}}
\renewcommand{\arraystretch}{1.2}
\begin{center}
\begin{tabular}{c| c c c}
\hline
\hline
Transients & $\rho_0$ ($\rm Gpc^{-3}yr^{-1}$) & $L_{\rm min}$ ($\rm erg\,s^{-1}$) & References  \\
\hline
SN SBO & \ratebpl{}  &$ 5\times 10^{42}$ & This work(BPL) \\

LL-GRB & $160^{+98}_{-65} $  & $5\times 10^{46}$ & (1)\tablenotemark{a} \\

HL-GRB & $2.4^{+0.3}_{-0.3}$   &  $3\times 10^{49}$ & (1)\\

SGRB & $4.2^{+1.3}_{-1.0} $  & $7\times 10^{49}$ & (1)\\

Thermal TDE & $5.0^{+2.3}_{-1.6}\times 10^3$  & $10^{43}$ & (1)\\

Jetted TDE & $0.03^{+0.04}_{-0.02} $  & $10^{48}$ & (1)\\

CDF-S XT2 & $1.4^{+3.3}_{-1.2} \times 10^4$  & $3\times 10^{45}$ & (2) \\

CDF-S XT1 & $1.1^{+2.5}_{-0.9} \times 10^4$  & $7 \times 10^{46}$ & (2) \\
\hline
\hline
\end{tabular}
\tablenotetext{a}{References (1) and (2) refer to \cite{sun2015} and \cite{sun2019}.}
\end{center}
\end{table*}

\subsection{$\log N -\log S$ distribution}
\begin{figure}
\centering
\includegraphics[scale=0.6]{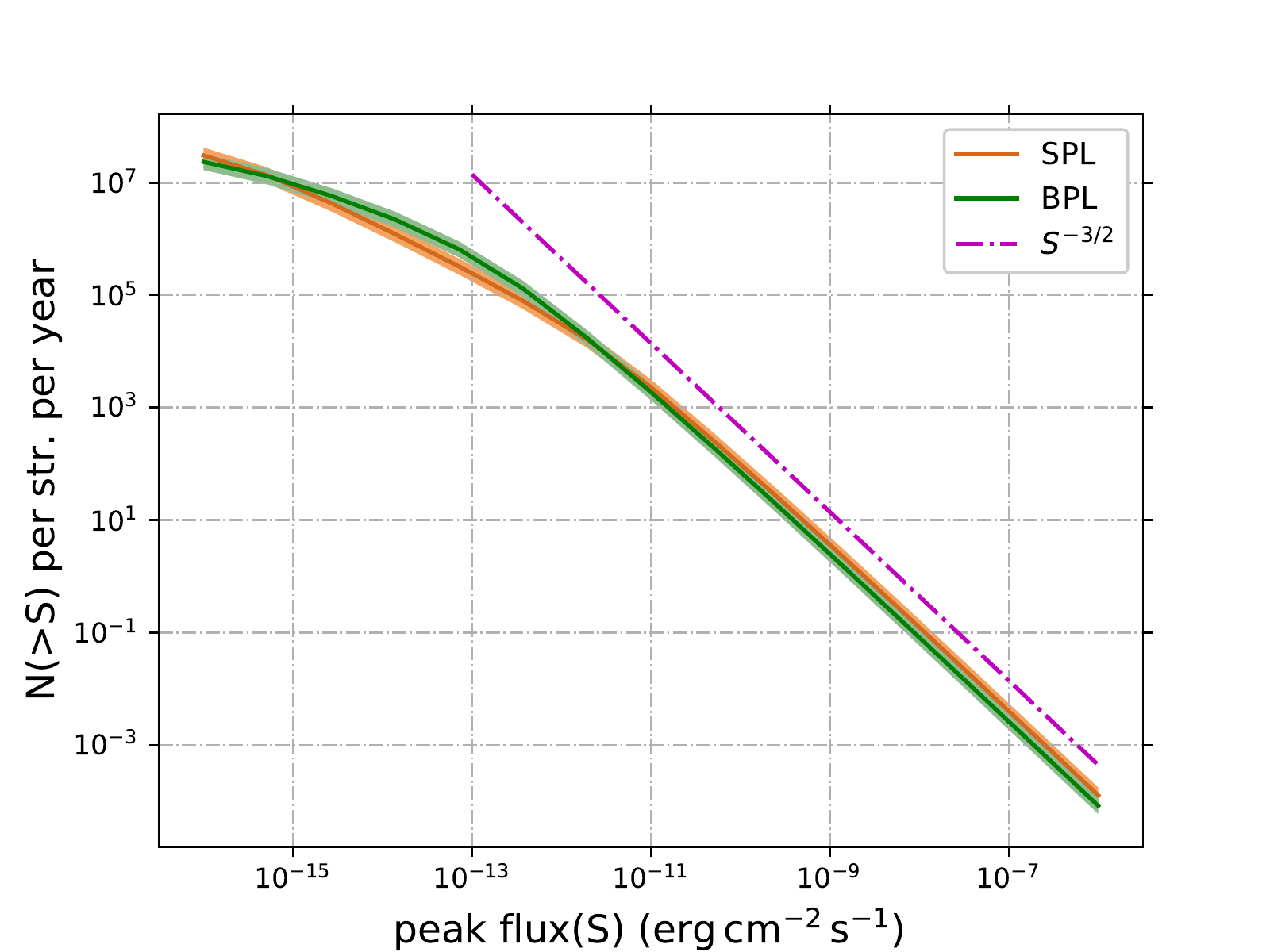} 
\caption{The $\log N -\log S$ distribution of SN SBOs for BPL (green solid line) and SPL (brown solid line) models, in units of per steradian per year. The shaded area gives the 1$\sigma$ uncertainty of local event rate density. The magenta dashed-dotted line shows an example of the relation for a population with uniform spatial distribution ($\approx S^{-3/2}$) for comparison.} 
\label{fig:lognlogs}
\end{figure}

We plot the SN SBO $\log N -\log S$ distribution in Figure \ref{fig:lognlogs}, which is calculated from the LF (both BPL and SPL) and the redshift evolution $f(z)$ from Equation \ref{eq:sfh}, where $N(>S)$  is the cumulative source counts with the peak flux above S. The numbers are given in units of per steradian per year. The peak luminosity is assumed to be in the range from $5 \times 10^{42}$ to $10^{47}$ $\rm erg\,s^{-1}$. 
The $\log N -\log S$ plot shows that when the peak flux is above $S>10^{-13}$ $\rm erg\,cm^{-2}s^{-1}$, the number generally declines following the $S^{-3/2}$ law (dashed-dotted line), as predicted for an Euclidean geometry. It flattens toward the lower flux end.
It should be pointed that the peak flux cannot be directly related to the sensitivity limit of the telescopes/surveys when one estimates the detection rate.

\section{Discussion and conclusions} \label{sec:DC}
\subsection{Comparison with CCSNe} \label{sec:ccsn}
We compare the local event rate density of SN SBOs with that of CCSNe inferred independently from optical SN searches. The local event rate density of CCSNe is given as $\sim 10^5$ $\rm Gpc^{-3}\,yr^{-1}$  \citep{smartt2009, liwd2011}. Among them, $\sim 59\%$ of the CCSNe are Type II-P SNe, which are expected to originate from explosions of red supergiants (RSGs), $\sim 1-3\%$ of CCSNe are Type II SNe similar to SN 1987A which are from the blue supergiant (BSG) collapse \citep{arnett1989, pastorello2005}, and $\sim 30\%$ of them are Type Ib/c SNe from the explosions of WR stars. It is suggested that the SN SBOs from RSGs (with a much larger radius and explosion energy), peak at a few eV and therefore are expected to be absorbed by the interstellar medium \citep{nakar2010}. In comparison, SN SBOs from the less massive BSGs are more easy to detect by soft X-rays detectors. 
The local event rate density of SN SBOs in this work is close to that of the Type II-P SNe but is around an order of magnitude higher than that of SN 1987A-like SNe. 
This can be interpreted as the estimated local event rate density being the total contribution by SN SBOs from the collapse of all three types of progenitors. In fact, two SN SBO candidates in AL20, XT 100424 and 151128, have been inferred with much larger radii $\geq 100R_{\odot}$ and softer temperature ($\sim 0.15$ $\rm keV$), which are more consistent with the predictions from RSGs. \cite{sapir2013} showed that even peaking at a lower energy band (like UV), SN SBOs from RSGs can emit a significant amount of photons in the soft X-rays, with the predicted luminosity ($\sim 10^{44}$ $\rm erg\,s^{-1}$) consistent with the observations of the two events. 

To perform a detailed comparison, we estimate the local event rate densities of each type of SN SBO separately. We follow the identification of AL20 and take those candidates that can be relatively well interpreted by one of the progenitor models, as summarized in Table \ref{tab:sample}. We derive the event rate densities following the above methodology. For the BSG SN SBO sample (XT 161028, XT 151219, XT 070618 and XT 060207), the event rate density is inferred as $2.1^{+0.8}_{-0.6} \times 10^3$ $\rm Gpc^{-3}yr^{-1}$ above the minimum luminosity $L_{\rm min, BSG} \sim 6\times 10^{44}$ $\rm erg\,s^{-1}$. For the RSG SN SBO sample (XT 100424 and XT 151128), one infers a rate of  $2.0 ^{+0.8}_{-0.6} \times 10^4$ $\rm Gpc^{-3}yr^{-1}$ above the minimum luminosity $L_{\rm min, RSG} \sim 5\times 10^{42}$ $\rm erg\,s^{-1}$. For WR SN SBOs, we take the less convincing candidates labeled BSG/WG for a rough estimate, i.e., both BSG and WR are possible origins. We derive a rate of  $1.7 ^{+0.7}_{-0.5} \times 10^4$ $\rm Gpc^{-3}yr^{-1}$ above the minimum luminosity $L_{\rm min, WR} \sim 10^{43}$ $\rm erg\,s^{-1}$. It is generally concluded that the event rate densities derived from the subsample and their relative ratios are consistent with the optically-inferred rates of CCSNe. One should be cautious that these estimates are affected by the classification of the sample. Furthermore, the event rate density inferred from RSG SN SBOs is around two times lower than that of Type II-P SNe. This discrepancy may be caused by different selection effects. For example, some X-ray RSG SN SBOs that are extremely soft may not be observed owing to severe Galactic absorption. There are also some SN SBOs peaking at UV band that will be missed by the X-ray population study, while the optical SNe observations do not suffer from such selection bias.

In addition, there seems to be a dip in the LF around the luminosity of ($\sim 10^{44}$ $\rm erg\,s^{-1}$) as shown in Figure \ref{fig:eventrate}, which is also consistent with the hypothesis that the LF is contributed by different populations of SN explosions. However, the current sample size is too small to draw a firm conclusion. 
The diversity of the SN SBOs from different progenitors can be confirmed by more events detected in the future.

\subsection{Prospect of the SN SBO detection with Einstein Probe}\label{sec:ep}
The Einstein Probe satellite is an all-sky monitor type space mission with a wide field X-ray telescope (WXT) on board \citep{yuan2018}. Adopting micropore optics technology for X-ray focusing imaging,  the WXT has a large FoV of 3600 square degrees and operates in the soft X-ray band of 0.5-4 $\rm keV$. The unprecedented Grasp (FoV $\times$ effective area) and soft X-ray bandpass make EP an ideal monitor to find transients like SN SBOs. The satellite plans to observe the nightside sky and scan half of the sky in every 3 orbits within 5 hours. We perform simulation of WXT all sky survey (Pan et al. in preparation) with the luminosity function, redshift evolution models in this work and use the light curves of the 12 candidates. The results show a detection of ten to several tens SN SBO events per year above the minimum peak flux  $\sim10^{-10}$ $\rm erg\,cm^{-2}s^{-1}$ for a detection with signal-to-noise ratio $S/N\,>\,5$, consistent with the prediction in Figure \ref{fig:lognlogs}. The majority of the secure detection is within the redshift of 0.1.  Large-area surveys in the optical band and spectroscopic follow-up observations will be necessary for identifying X-ray transients with an SN SBO origin. 

\subsection{Summary}
In this work, we provide a realistic method of deriving the LF and event rate density for fast X-ray transients, and apply it to the newly reported SN SBO sample discovered from \textit{XMM-Newton} archival data. We derive the maximum volume  ($V_{\rm max}$)  within which a source can be detected for a specific telescope or survey from simulations of more realistic detection of transient sources using the data sets of the real observations.  It is shown that the SN SBO LF can be a BPL with indices (at the 68$\%$ confidence level) of \bplslopef{} and \bplslopes{} before and after the break luminosity $\log (L_b/\rm erg\,s^{-1})=$ \bpllbreak{} or an SPL with index of \plslope{}. The local event rate densities of SN SBOs above $5 \times 10^{42}$ $\rm erg\,s^{-1}$ are \ratebpl{} and \ratepl{} $\rm Gpc^{-3}\,yr^{-1}$ for BPL and SPL models, respectively. 
We perform the joint LF for SN SBOs and LL-GRBs and the SPL indices are similar to that of the SPL fit of SN SBO. It should be noted that the results are affected by the systematic uncertainties due to the binning choices and the possibility of misclassified SBOs. Though the LFs and event rate densities derived from different cases of bin width and from the alternative subsample are moderately different, they are mostly within each other's error range. 

We interpret the local event rate density as being the total contribution by SN SBOs from the collapse of all three types of progenitors (i.e., RSGs, WRs and BSGs).
The LF of SN SBOs in the low luminosity range is mainly contributed by the explosion of RSGs and WRs. The BSG SN SBOs tend to have higher peak luminosities and an event rate density an order of magnitude lower than that of the former two.
We also present a $\log N -\log S$ distribution of SN SBOs based on the LF results. It is expected that a wide field X-ray telescope like EP can detect ten to a few tens of SN SBOs per year. 

\acknowledgments
We thank an anonymous referee for constructive comments and suggestions. We are grateful to Haonan Yang and Wenda Zhang for helpful discussions. This work is supported by the Strategic Pioneer Program on Space Science, Chinese Academy of Sciences, Grant No. XDA15052100. H.S. is supported by National Natural Science Foundation of China (No. 12103065). H.Y.L. is supported by China Postdoctoral Science Foundation (2021M693203) and National Natural Science Foundation of China (No. 12103061). H.-W.P. is supported by National Natural Science Foundation of China (Nos. 11803047, U1838202, 11873054).
Z.L. is supported by the Natural Science Foundation of China (No. 11773003, U1931201). This research has made use of Astropy, a community-developed core Python package for Astronomy \citep{astropy}, Matplotlib \citep{matplotlib} and \textit{Uncertainties}: a Python package for calculations with uncertainties, Eric O. LEBIGOT. This research has used the calibration files provided by the High Energy Astrophysics Science Archive Research Center Online Service, provided by the NASA/Goddard Space
Flight Center.

\bibliography{sbo_ms}{}
\bibliographystyle{aasjournal}

\end{document}